\documentstyle[12pt,fleqn]{article}
\newtheorem{prop}{Proposition}[section]
\newtheorem{lem}[prop]{Lemma}
\newtheorem{cor}[prop]{Corollary}

\newcommand{\Ibb}[1]{ {\rm I\ifmmode\mkern
            -3.6mu\else\kern -.2em\fi#1}}
\newcommand{\ibb}[1]{\leavevmode\hbox{\kern.3em\vrule
     height 1.2ex depth -.3ex width .2pt\kern-.3em\rm#1}}
\newcommand{\Cx}{{\ibb C}}

\newcommand{\Rl}{{\Ibb R}}

\newcommand{\ed}{\mbox{\rm d}}
\newcommand{\id}{\mbox{\rm id}}
\newcommand{\kro}[2]{\mbox{$\delta^{#1}{\!}_#2$}}

\newcommand{\zz}[2]{\mbox{$z^{\,#1}{}_#2$}}
\newcommand{\tz}[2]{\mbox{$\theta^{\,#1}{}_#2$}}
\newcommand{\htz}[2]{\mbox{$\hat{\theta}^{\,#1}{}_#2$}}
\newcommand{\un}{{\bf 1}}
\newcommand{\de}{{\cal D}}
\newcommand{\ide}{{\cal D}^{-1}}

\newcommand{\iz}{$z^{\,i}{}_j \ $}
\newcommand{\ith}{$\theta^{\,i}{}_j \ $}
\newcommand{\ei}{$\bf 1 \ $}
\newcommand{\ca}{$\cal A \ $}
\newcommand{\Ga}{$\Gamma \ $}

\newcommand{\lco}{\Delta_{\cal L}}
\newcommand{\rco}{\Delta_{\cal R}}

\newcommand{\mca}{{\cal A}}
\newcommand{\Gal}{\mbox{$_{inv}\Gamma$}}

\newcommand{\GL}{GL$_q(3,\Cx)$}
\newcommand{\SL}{SL$_q(3,\Cx)$}

\newcommand{\Trq}{\mbox{\rm Tr}_q}

\newcommand{\zwi}[3]{\mbox{$#1^{\,#2}{}_#3$}}
\newcommand{\zwia}[3]{\mbox{$#1_{#2}{\,}^#3$}}
\newcommand{\vii}[5]{\renewcommand{\arraystretch}{0.45}
\mbox{$#1\!\! \begin{array}{l}\scriptstyle #2\,
#4\\ \scriptstyle\ #3\, #5\end{array}$}\renewcommand{\arraystretch}{1}}
\newcommand{\viia}[5]{\mbox{$#1^{\,#2}{}_{#3 \, #4}{\,}^#5$}}
\newcommand{\viib}[5]{\mbox{$#1_{#2}{\,}^{#3 \, #4}{\,}_#5$}}

\newcommand{\lpm}{\ell^{\pm}{}}
\newcommand{\lmp}{\ell^{\mp}{}}

\newcommand{\th}[1]{{\rm #1}}

\begin{document}
\thispagestyle{empty}
\vspace*{3cm}
\centerline{\LARGE \bf All bicovariant differential calculi}
\vskip.5cm
\centerline{\LARGE \bf on GL$_q(3,\Cx)$ and SL$_q(3,\Cx)$}
\vskip2.5cm
\centerline{\large \bf Klaus Bresser}
\vskip.3cm
\centerline{Institut f\"ur Theoretische Physik}
\centerline{der Universit\"at G\"ottingen}
\centerline{Bunsenstr. 9, D-37073 G\"ottingen}

\vspace{3cm}
\begin{abstract} \noindent
All bicovariant first order differential calculi on the quantum
group \GL\ are determined. There are two distinct one-parameter families
of calculi. In terms of a suitable basis of 1-forms the commutation
relations can be expressed with the help of the $R$-matrix of \GL.
Some calculi induce bicovariant differential calculi on \SL\ and on
real forms of \GL. For generic deformation parameter $q$ there are
six calculi on \SL, on SU$_q(3)$\ there are only two.
The classical limit $q \to 1$ of bicovariant calculi on \SL\ is not
the ordinary calculus on SL$(3,\Cx)$. One obtains a deformation of it
which involves the Cartan-Killing metric.
\end{abstract}
\newpage

\section{Introduction}

In recent years `non-commutative geometry' (see \cite{Co86,Coq90} for
some aspects of it) appeared as a new branch of geometry and a new
framework for physical model building. It has its origin in the basic
observation that a manifold (respectively, a topological space) is
completely characterized by the algebra of functions on it, viewed as
an abstract commutative ($C^\ast$-) algebra. Geometrical concepts
can be understood as algebraic structures on this algebra and then
generalized to non-commutative algebras (for which there is no longer
an underlying topological space).

In differential geometry an important role is played by Lie groups
which correspond to commutative Hopf algebras \cite{Abe80,Sw69}.
`Quantum groups' are non-commutative Hopf algebras. Examples are
obtained as deformations of classical groups (as Hopf algebras)
\cite{Dri86,Wo87,Maj90,Tj92}. In particular, they provide us with
new symmetry concepts which are of relevance, in particular, in the
context of conformal field theories and quantum integrable models.

Differential geometry of Lie groups (and their coset spaces)
enters the mathematical modelling of physical theories. In particular,
this is the case for classical gauge theories formulated in terms of
connections on principal fiber bundles, and for Kaluza-Klein theories.
First steps have been made to generalize the corresponding notions to
the realm of non-commutative geometry (see \cite{BM93,BM93a,Pf92}, for
example). There is some hope to obtain interesting `deformations'
of physical models in this way, in particular for elementary particle
physics and gravitation.

A central part of such a program is to develop differential calculus
on quantum qroups. This has been done by Woronowicz \cite{Wo89}. He
introduced the notion of bicovariance as a natural condition to reduce
the number of possible differential algebras associated with a given
quantum group. In the meantime a large number of papers appeared dealing
with examples of bicovariant differential calculi on special (classes
of) quantum groups (see \cite{MHR93} for an extensive list of
references). However, one would like to have a complete description of
all possible bicovariant differential calculi on certain quantum groups
rather than just a collection of examples.
For the two-parameter quantum group GL$_{p,q}(2,\Cx)$ and related
subgroups this was achieved in \cite{MH92} and \cite{MHR93}.
We used similar methods to determine all bicovariant (first order)
differential calculi on \GL\ and \SL.\footnote{These results were
communicated at the spring meeting of the Deutsche Physikalische
Gesellschaft in Hamburg, March 1994.}
Examples of bicovariant differential calculi on \GL\ have already
been presented in \cite{CA92}.

The classical limit $q \rightarrow 1$ leads to a Hopf algebraic
description of the Lie groups GL$(3,\Cx)$ and SL$(3,\Cx)$. One might
expect the usual differential geometry of these groups to be
recovered in this limit. However, for $q \rightarrow 1$ we obtain
an interesting deformation of the ordinary differential calculus on
SL$(3,\Cx)$ (see also \cite{MHR93} for the case of SL$(2,\Cx)$). Functions on
the group no longer commute with 1-forms, the commutation relations
involve the Cartan-Killing metric. This observation may be taken as a
starting point for further investigations aiming at the notion of a
`quantum group metric'.

Section 2 recalls the notions of differential calculus and
bicovariance on quantum groups. In section 3 we briefly review the
Hopf algebraic structure of \GL. The central part of our work is
section 4 which deals with the determination of all bicovariant
differential calculi on \GL\ and a discussion of some of their
properties. In section 5 we turn to the investigation of bicovariant
calculi on quantum subgroups of \GL.
Section 6 is devoted to the classical limit of bicovariant differential
calculi on \GL\ and \SL. Finally, in section 7 we relate our results to
the work of other authors and try to give a perspective for further
studies.

\section{Differential calculus on quantum groups}
\setcounter{equation}{0}
We first recall the definition of a (first order) differential
calculus on an associative algebra $\cal A$ and specify later
to the case of a Hopf algebra (respectively, a quantum group)
\cite{Wo89}.

{\bf Definition.} Let \ca be an associative unital algebra.  An
$\cal A$-bimodule $\Gamma$ together with a linear map $\ed : {\cal A}
\longrightarrow \Gamma$ is called first order differential calculus
over $\cal A$ iff
\begin{itemize}
   \item[(1)] $\ed (ab) = (\ed a)b + a(\ed b) \quad$ for all $a,b \in {\cal
A},$
   \item[(2)] d$\cal A$ generates $\Gamma$ as left $\cal A$-module.
\end{itemize}
Two first order differential calculi ($\Gamma, \ed$) and
($\tilde{\Gamma}, \tilde{\ed}$) over $\cal A$ are said to be equivalent
iff there exists a bimodule isomorphism $\zeta : \Gamma
\longrightarrow \tilde{\Gamma}$  with $\tilde{\ed} = \zeta \circ
\ed$.
This definition generalizes the classical notion of first order
differential forms. We will therefore call the elements of \Ga
1-forms.

Let us now turn to the case of a Hopf algebra.
Besides the multiplication and the unit element a quantum group
carries the following additional structure:
\begin{equation}
\begin{array}{ll}
\Delta : \mca \longrightarrow \mca \otimes \mca & \mbox{(coproduct)} \\[1ex]
\varepsilon : \mca \longrightarrow \Cx & \mbox{(counit)} \\[1ex]
S : \mca \longrightarrow \mca & \mbox{(antipode)}
\end{array}
\end{equation}
The first two maps are algebra homomorphisms, the third is an
algebra antihomomorphism. These maps have to fulfil certain axioms
which we need not recall here (cf \cite{Abe80,Sw69,Maj90}).
In the commutative case they encode the group structure of the
underlying group manifold in the algebraic structure of the algebra of
functions on the group. In particular, the coproduct translates the
group multiplication and can be used to reformulate the left and right
action of the group on itself.
One may now ask whether there are corresponding generalizations of the
induced actions of the group on differential forms. This leads to the
notion of bicovariance which is briefly recalled in the sequel.

{\bf Definition.} Let \ca be a Hopf algebra with unit element \ei. A
first order differential calculus (\Ga, d) over \ca is called
bicovariant iff there are linear maps $\Delta_{\cal L}: \Gamma
\rightarrow {\cal A} \otimes \Gamma$ and
$\Delta_{\cal R}: \Gamma \rightarrow \Gamma \otimes {\cal A}$, which
are called left and right coactions, such that
\begin{eqnarray}  \label{lcod}
   \Delta_{\cal L}(a \ed b) & = & \Delta(a)(\id \otimes \ed)\Delta(b)
\\[1ex]  \label{rcod}
   \Delta_{\cal R}(a \ed b) & = & \Delta(a)(\ed \otimes \id)\Delta(b)
   \ .
\end{eqnarray}

An element $\omega \in \Gamma$ is said to be left-/right-invariant
iff
\begin{eqnarray} \label{linv}
   \lco (\omega) & = & \un \otimes \omega  \\[1ex] \label{rinv}
   \rco (\omega) & = & \omega \otimes \un
\end{eqnarray}
respectively.
$\omega$ \ is called bi-invariant iff (\ref{linv}) and (\ref{rinv})
hold simultaneously.

A bicovariant differential calculus is a special case of a structure
called bicovariant bimodule, which is by definition an \ca-bimodule
\Ga together with linear maps $\Delta_{\cal L}: \Gamma \rightarrow {\cal A}
\otimes \Gamma$ and $\Delta_{\cal R}: \Gamma \rightarrow \Gamma \otimes
{\cal A}$ satisfying
\begin{displaymath}
   \begin{array}{r@{\: = \:}l}
   \lco(a \varrho b) & \Delta(a) \lco(\varrho) \Delta(b) \\[1ex]
   \rco(a \varrho b) & \Delta(a) \rco(\varrho) \Delta(b)
   \end{array}
\end{displaymath}
\begin{displaymath}
   \begin{array}{r@{\: = \:}l}
   (\id \otimes \lco) \circ \lco & (\Delta \otimes \id) \circ \lco
\\[1ex]
   (\rco \otimes \id) \circ \rco & (\id \otimes \Delta) \circ \rco
   \end{array}
\end{displaymath}
\begin{displaymath}
   \begin{array}{r@{\: = \:}l}
   (\varepsilon \otimes \id) \circ \lco(\varrho) & \varrho \\[1ex]
   (\id \otimes \varepsilon) \circ \rco(\varrho) & \varrho
   \end{array}
\end{displaymath}
and
\begin{displaymath}
   (\id \otimes \rco) \circ \lco = (\lco \otimes \id) \circ \rco \ .
\end{displaymath}
For $\lco$\ and $\rco$\ given by (\ref{lcod}) and (\ref{rcod}) these
identities are satisfied. It turns out that the whole structure of a
bicovariant bimodule \Ga can be conveniently described by its left-
(or right-) invariant elements.
We introduce the left and right convolution products, defined for $f
\in \mca^{\prime} = \mbox{Hom}(\mca, \Cx)$ and $a \in \mca$ by
\begin{eqnarray}
   f \ast a & = & (\id \otimes f) \Delta (a) \\[1ex]
   a \ast f & = & (f \otimes \id) \Delta (a)
\end{eqnarray}
and recall some results from \cite{Wo89}.

\begin{prop} \label{bicostru} Let $(\Gamma, \ed)$\ be a bicovariant
bimodule over the Hopf algebra \ca. The set of all left-invariant
elements of $\Gamma$, called $_{inv}\Gamma$, is a linear subspace of
$\Gamma$. Let $\{\omega^{\,I}\}_{I \in \cal I} $ be a basis of
$_{inv}\Gamma$. Then:
\begin{itemize}
   \item[\th{(1)}] Any $\rho \in \Gamma \ $can uniquely be written as
   $\rho = a_{I} \, \omega^{\,I}$ \ with $a_{I} \in \mca$.
   \item[\th{(2)}] There exist linear functionals $f^{\,I}{}_J \in
              \mca^{\prime} $ such that
\begin{equation} \label{comw}
   \omega^{\,I}a = (f^{\,I}{}_J \ast a) \omega^{\,J} \quad \forall I
   \in {\cal I} \ \ \forall a \in \mca.
\end{equation}
The functionals are uniquely determined by \th{(\ref{comw})} and fulfil
the relations
\begin{eqnarray} \label{re-prop}
   f^{\,I}{}_J (ab) & = & f^{\,I}{}_K (a) f^{\,K}{}_J (b) \\[1ex]
   \label{re-prop1}
   f^{\,I}{}_J (\un) & = & \delta^{\,I}{}_J \ .
\end{eqnarray}
   \item[\th{(3)}] The right coaction on the basis $\{\omega^{\,I}\}_{I
              \in \cal I} $ is given by
\begin{equation} \label{rcoom}
   \rco (\omega^{\,I}) = \omega^{\,J} \otimes M_J{}^{I}
\end{equation}
with $M_J{}^{I} \in \mca$ satisfying
\begin{eqnarray} \label{ad-re}
   \Delta(M_I{}^{J}) & = & M_I{}^{K} \otimes M_K{}^{J} \\
   \label{ad-re1}
   \epsilon(M_I{}^{J}) & = & \delta_I{}^{J} \ .
\end{eqnarray}
   \item[\th{(4)}] Bicovariance implies
\begin{equation} \label{bico-con}
   M_I{}^{J} (a \ast f^{\,I}{}_K) = (f^{\,J}{}_I \ast a) M_K{}^{I}
   \quad \forall \, a \in \mca \ \ \forall \, J, K \in {\cal I}.
\end{equation}
\end{itemize}
\end{prop}

In this short exposition we will not consider the higher order
differential calculus. We only mention that every bicovariant first
order differential calculus admits an extension to a differential
algebra containing forms of arbitrary order (cf. \cite{Wo89,Br93}).

\section{The quantum group GL$_q(3,\Cx)$}
\setcounter{equation}{0}

Deformations of Lie groups can be obtained by introducing a
non-commutative multiplication structure on the related Hopf algebra.
This usually involves deformation parameters. Corresponding
multi-parameter deformations of (the algebra of functions on) the
general linear groups are known (cf \cite{Ma89,ASchT91,Schirr91}).
Examples of differential calculi have been constructed on some of them
\cite{Mal90,Ma91,MH92}. Here we concentrate on the standard
one-parameter deformation of the algebra of functions on GL$(3,\Cx)$
\cite{FRT}. This is the algebra
$\mca := \mbox{Fun}_q(\mbox{GL}(3,\Cx))$ generated by
\begin{itemize}
   \item[(a)] nine noncommuting entities \iz , $i,j = 1,2,3$, which
we arrange as a matrix $Z = (z^{\,i}{}_j)$. Their commutation relations
are
\begin{equation}   \label{explRZZ}
\begin{array}{cccrcl}
j<k: &  & \quad & z^{\,i}{}_j\ z^{\,i}{}_k & = & q\ z^{\,i}{}_k\
                                                 z^{\,i}{}_j  \\
[1ex]
i<k: &  & \quad & z^{\,i}{}_j\ z^{\,k}{}_j & = & q\ z^{\,k}{}_j\
                                                 z^{\,i}{}_j  \\
[1ex]
i<k,\ & j>l: & \quad & z^{\,i}{}_j\ z^{\,k}{}_l & = & z^{\,k}{}_l\
                                                 z^{\,i}{}_j   \\
[1ex]
i<k,\ & j<l: & \quad & z^{\,i}{}_j\ z^{\,k}{}_l & = & z^{\,k}{}_l\
z^{\,i}{}_j+(q-q^{-1})\ z^{\,k}{}_j\ z^{\,i}{}_l  \ .
\end{array}
\end{equation}
For $q \to 1 $ all the matrix elements of $Z$ commute
with each other (classical limit). Sometimes it is convenient to
treat the indices $^{i}{}_j \ $ of \iz as `composite indices'
taking values 1,...,9 (via $^{1}{}_1 \rightarrow 1,\  ^{1}{}_2
\rightarrow 2,\  ^{1}{}_3 \rightarrow 3,\  ^{2}{}_1 \rightarrow 4,\
$ etc.).
   \item[(b)] the unit \ei \ and the inverse $\ide$ of the quantum
determinant
\setlength{\mathindent}{0.5cm}
\begin{equation}
   {\cal D} = z^1z^5z^9 + q^2 z^2z^6z^7 + q^2 z^3z^4z^8 - q z^1z^6z^8 -
              q^3 z^3z^5z^7 - q z^2z^4z^9 \ ,
\end{equation}
\setlength{\mathindent}{1cm}
which is central in \ca.
\end{itemize}

This non-commutative algebra can be endowed with a coproduct, counit and
antipode in the following way:
\begin{equation}
\begin{array}{r@{\: = \:}l@{\qquad\ \ }r@{\: = \:}l}
\Delta(z^{\,i}{}_j) & z^{\,i}{}_k \otimes z^{\,k}{}_j  &
\Delta({\bf 1}) & {\bf 1} \otimes {\bf 1} \\ [1ex]
\varepsilon(z^{\,i}{}_j) &  \delta^{\,i}{}_j &
\varepsilon({\bf 1}) & 1   \\[1ex]
S(z^{\,i}{}_j) & (S(Z))^{\,i}{}_j & S({\bf 1}) &
{\bf 1}
\end{array}
\end{equation}
where the summation convention is used and the matrix $S(Z)$ is given by
\setlength{\mathindent}{0.5cm}
\begin{equation}
S(Z) = {\cal D}^{-1} \left( \begin{array}{ccc}
z^5z^9-qz^6z^8 & -q^{-1}z^2z^9+z^3z^8 & q^{-2}z^2z^6-q^{-1}z^3z^5 \\
-qz^4z^9+q^2z^6z^7 & z^1z^9-qz^3z^7 & -q^{-1}z^1z^6+z^3z^4 \\
q^2z^4z^8-q^3z^5z^7 & -qz^1z^8+q^2z^2z^7 & z^1z^5-qz^2z^4
\end{array} \right) .
\end{equation}
\setlength{\mathindent}{1cm}

$({\cal A}, \cdot, {\bf 1}, \Delta, \epsilon, S)$ then constitutes a
Hopf algebra which may formally be regarded as an algebra of
`functions' on some (fictitious) space \GL.

{\bf Remark.} In a similar way one obtains the Hopf algebra
$\mbox{Fun}_q(\mbox{GL}(n,\Cx))$ using $n^2$ generators $Z = (\zz i j)$.
Let $Z_1 = Z \otimes I, Z_2 = I \otimes Z \ $where $I$ is the $n \times n$
unit matrix. The relations (\ref{explRZZ}) can be written
in compact form
\begin{equation} \label{RZZ}
   R_{12}Z_1Z_2 = Z_2Z_1R_{12}
\end{equation}
with the help of the nonsingular complex matrix $R \in \mbox{M}(n^2,\Cx)$
\begin{equation}  \label{R-mat}
   R = \sum_{i,j=1}^{n} q^{\delta^i{}_j} \, \zwia e i i \otimes \zwia e j j
+ (q-q^{-1})\sum_{i,j=1 \atop i>j}^{n}\zwia e i j \otimes \zwia e j i \
,\quad q \in \Cx^{\ast}
\end{equation}
where the matrices $\zwia e i j$ are defined by
$(\zwia e i j)^{\,k}{}_l = \kro k i \kro l j$.
In this form the associativity of \ca is conveniently expressed by
the quantum Yang-Baxter equation
\begin{equation} \label{QYBE}
   R_{12}R_{13}R_{23} = R_{23}R_{13}R_{12} \ .
\end{equation}
The antipode of \ca is invertible. Defining a diagonal matrix
$D = \mbox{diag}(1, q^2,\dots,q^{2(n-1)})$\ one has
\begin{equation} \label{invant}
   S^{-1}(Z) = D^{-1} S(Z) D \ .
\end{equation}

\section{Bicovariant differential calculus on GL$_q(3,\Cx)$}
\setcounter{equation}{0}

Let $(\Gamma, \mbox{d})$ be a first order differential calculus over
$\mca := \mbox{Fun}_q(\mbox{GL}(3,\Cx))$. $\Gamma$ is generated by the
differentials $\mbox{d}z^{\,i}{}_j \ (i,j = 1,2,3)$ as an
$\cal A$-bimodule. The differentials of the other generators are
obtained using the Leibniz rule:
\begin{eqnarray}
   \ed{\bf 1} & = & 0 \ ,\\ [1ex]
   \ed{\cal D}^{-1} & = & -{\cal D}^{-1}(\ed{\cal D}){\cal D}^{-1}
   \ .
\end{eqnarray}

To mimic the case of (commutative) differential geometry it is natural
to require that $\Gamma$ is generated by $\mbox{d}z^{\,i}{}_j \
(i,j = 1,2,3)$ as a left $\cal A$-module. This asssumption will be
made in the sequel.

Now we proceed along the lines of \cite{MH92} with
emphasis on the fundamental results of \cite{Wo89}.

\subsection{The left-invariant Maurer-Cartan 1-forms}

In order to determine the most general commutation relations of
elements of \Ga with elements of \ca we use a convenient basis of
$\Gamma$.
It consists of the quantum analogues of the Maurer-Cartan 1-forms
defined by
\begin{equation}    \label{MCforms}
   \theta^{\,i}{}_j = S(z^{\,i}{}_k) \,\mbox{d}z^{\,k}{}_j \ .
\end{equation}

The relevant properties of these 1-forms are summarized in

\begin{lem} \th{(1)} The 1-forms \ith are left-invariant, i.e.
\begin{equation}
   \Delta_{\cal L} (\theta^{\,i}{}_j) = {\bf 1} \otimes \theta^{\,i}{}_j
   \ .
\end{equation}

\th{(2)} The set ${\cal B} := \{ \theta^{\,i}{}_j \mid i,j = 1,2,3 \}$ is a
basis of $_{inv}\Gamma$ as a $\Cx$-vectorspace.

\th{(3)} For the right coaction on \ith, one finds
\begin{equation} \label{rcth}
   \Delta_{\cal R}(\theta^{\,i}{}_j) = \theta^{\,m}{}_n \otimes
M_m{}^{n \, i}{}_j \ , \quad M_m{}^{n \, i}{}_j := S(z^{\,i}{}_m) z^{\,n}{}_j
\ \in {\cal A} \ .
\end{equation}

\end{lem}

By forming composite indices from the matrix indices (see section 3)
one obtains (\ref{rcoom}) with $M_J{}^{I}$ satisfying
(\ref{ad-re}) and (\ref{ad-re1}).

{\bf Remark.} Using (\ref{invant}) one can verify the identity
$\sum_i q^{-2i} S(\zz il) \zz k i = q^{-2k} \kro k l$. This shows that
$\Trq\, \theta = \sum_i q^{6-2i} \tz i i$\ is a bi-invariant element
of \Ga.

\subsection{Structure of the commutation relations}

Since the Maurer-Cartan 1-forms $\theta^{\,i}{}_j$ form a basis
of the space of all left-invariant 1-forms $_{inv}\Gamma$ we
have uniquely determined linear functionals $f^{\,I}{}_J \in {\cal
A}^{\prime},\ 1 \le I,J \le 9,$ such that
\begin{equation} \label{comth}
   \theta^{\,I}a = (f^{\,I}{}_J \ast a) \theta^{\,J} = ((\mbox{id}
\otimes f^{\,I}{}_J) \circ \Delta(a)) \theta^{\,J}
\end{equation}
for all $a \in \cal A$\ (Proposition \ref{bicostru}). Because of
(\ref{re-prop}) and (\ref{re-prop1}) these functionals provide us with a
representation ${\cal F}:{\cal A} \longrightarrow \mbox{M}(9,\Cx)$.
The `fundamental matrices'
\begin{equation}
{\cal F}(z^{\,i}{}_j) = (f^{\,I}{}_J(z^{\,i}{}_j))_{I,J=1,\ldots,9}
\ .
\end{equation}
completely and uniquely specify the first order differential calculus
(using the equivalence definition of section 2.1).

There are restrictive conditions which a set of matrices has to satisfy
in order to be the fundamental matrices of a bicovariant differential
calculus on $\cal A$:
\begin{itemize}
   \item[(1)] {\bf Consistency with the commutation relations of $\cal
A$:}

By differentiating the commutation relations (\ref{RZZ}) one
obtains
\begin{displaymath}
   0 = \ed(RZ_1Z_2 - Z_2Z_1R) = R \ed Z_1 Z_2 + RZ_1 \ed Z_2 - \ed Z_2
Z_1R - Z_2 \ed Z_1 R \ .
\end{displaymath}
After convertion of the differentials into Maurer-Cartan
forms and commuting all algebra elements to the left we get
conditions for the values $f^{\,I}{}_J(z^{\,i}{}_j)$ of the
functionals $f^{\,I}{}_J$.

   \item[(2)] {\bf Bicovariance conditions (\ref{bico-con}):}

Inserting the algebra generators \iz in (\ref{bico-con}) and using
(\ref{comth}) further conditions are obtained for the values
$f^{\,I}{}_J(z^{\,i}{}_j)$.

   \item[(3)] {\bf Representation properties of the functionals
$f^{\,I}{}_J$:}

Acting with $\cal F$ on the commutation relations (\ref{RZZ}) and
using the representation property of $\cal F$ leads to further
equations for the matrices ${\cal F}(z^{\,i}{}_j)$. These are
nonlinear equations, in general. Furthermore, ${\cal F}({\cal D})$
has to be invertible in $\mbox{M}(9,\Cx)$.
\end{itemize}

Using the conditions (1)--(3) one can derive the most general set
of matrices ${\cal F}(z^{\,i}{}_j)$ which determines a bicovariant
differential calculus. For this purpose we used the computer algebra
software {\sc Reduce}. It is convenient to solve the equations
resulting from (1) and (2) first because they are linear in the
matrix elements. Using finally the equations resulting from condition
(3) we are led to the following results.

\subsection{Results} \label{subres}

\begin{prop} Let $q \in \Cx \setminus \{0,\pm 1,\pm i\}$. Then all
bicovariant differential calculi on \th{GL$_q(3,\Cx)$}\ are contained
in two disjoint one-parameter families of calculi denoted by
$\Gamma_{\nu}(t),\ \nu = 1,2$ where
\begin{displaymath}
t \in \Cx \setminus \{0\}, \quad (q^6+q^4+1)t-(q^6+q^4+q^2) \neq 0
\end{displaymath}
in the first and
\begin{displaymath}
t \in \Cx \setminus \{0\}, \quad (q^6+q^2+1)t-(q^4+q^2+1) \neq 0
\end{displaymath}
in the second case.
The calculi $\Gamma_{\nu}(t)$\ and $\Gamma_{\nu^{\prime}}(t^{\prime})$ are
equivalent if and only if $\nu = \nu^{\prime}$\ and $t = t^{\prime}$.
\end{prop}

{\bf Remark.}
The calculi can be described explicitly in terms of their fundamental
matrices ${\cal F}(z^{\,i}{}_j) \ (i,j = 1,2,3)$ which depend on $q$ and
the extra parameter $t$. The rather lengthy expressions can be found
in \cite{Br94}. For the exceptional values $q = \pm 1,\pm i \ $ there
may be further calculi.

Now one can calculate the commutation relations of the generators of
\ca and their differentials from the commutation relations involving
the Maurer-Cartan 1-forms. These calculations have also been carried
out with the help of {\sc Reduce}.

\begin{cor} Let $q \in \Cx \setminus \{0,\pm 1,\pm i\}$. The
bicovariant differential calculi $\Gamma_1(t)$\ on
\th{GL$_q(3,\Cx)$}\ are given by
\setlength{\mathindent}{0.25cm}
\begin{equation}  \label{dzz1}
\begin{array}{rclc}
(\ed z^{\,i}{}_j) z^{\,i}{}_j & = & (\frac{t}{q^2}+t-1) z^{\,i}{}_j
\ed z^{\,i}{}_j + \sigma z^{\,i}{}_j z^{\,i}{}_j \Trq \theta &  \\[1ex]
(\ed z^{\,i}{}_j) z^{\,i}{}_l & = & \frac{t}{q} z^{\,i}{}_l
\ed z^{\,i}{}_j + (t-1) z^{\,i}{}_j \ed z^{\,i}{}_l + \sigma z^{\,i}{}_j
z^{\,i}{}_l \Trq \theta & j < l \\[1ex]
(\ed z^{\,i}{}_j) z^{\,k}{}_j & = & \frac{t}{q} z^{\,k}{}_j
\ed z^{\,i}{}_j + (t-1) z^{\,i}{}_j \ed z^{\,k}{}_j + \sigma z^{\,i}{}_j
z^{\,k}{}_j \Trq \theta & i < k \\[1ex]
(\ed z^{\,i}{}_j) z^{\,k}{}_l & = & t z^{\,k}{}_l
\ed z^{\,i}{}_j + (t-1) z^{\,i}{}_j \ed z^{\,k}{}_l + \sigma z^{\,i}{}_j
z^{\,k}{}_l \Trq \theta \\[1ex]
& & - \beta (z^{\,i}{}_j z^{\,k}{}_l - q z^{\,i}{}_l z^{\,k}{}_j) \Trq
\theta & i < k,j < l \\[1ex]
(\ed z^{\,i}{}_j) z^{\,k}{}_l & = & t z^{\,k}{}_l
\ed z^{\,i}{}_j + (t-1) z^{\,i}{}_j \ed z^{\,k}{}_l + \sigma z^{\,i}{}_j
z^{\,k}{}_l \Trq \theta \\[1ex]
& & - t(q-\frac{1}{q}) z^{\,k}{}_j \ed z^{\,i}{}_l + q \beta
(z^{\,i}{}_l z^{\,k}{}_j - q z^{\,i}{}_j z^{\,k}{}_l) \Trq \theta &
i < k,j > l \\[1ex]
\end{array}
\end{equation}
with $t \in \Cx \setminus \{0\},\ (q^6+q^4+1)t-(q^6+q^4+q^2) \neq 0$. The
second family of calculi $\Gamma_2(t)$\ is determined by
\begin{equation}  \label{dzz2}
\begin{array}{rclc}
(\ed z^{\,i}{}_j) z^{\,i}{}_j & = & (t q^2 + t - 1) z^{\,i}{}_j
\ed z^{\,i}{}_j + \hat{\sigma} z^{\,i}{}_j z^{\,i}{}_j \Trq \theta &  \\[1ex]
(\ed z^{\,i}{}_j) z^{\,i}{}_l & = & tq z^{\,i}{}_l
\ed z^{\,i}{}_j + (tq^2-1) z^{\,i}{}_j \ed z^{\,i}{}_l + \hat{\sigma}
z^{\,i}{}_j
z^{\,i}{}_l \Trq \theta & j < l \\[1ex]
(\ed z^{\,i}{}_j) z^{\,k}{}_j & = & tq z^{\,k}{}_j
\ed z^{\,i}{}_j + (tq^2-1) z^{\,i}{}_j \ed z^{\,k}{}_j + \hat{\sigma}
z^{\,i}{}_j
z^{\,k}{}_j \Trq \theta & i < k \\[1ex]
(\ed z^{\,i}{}_j) z^{\,k}{}_l & = & t z^{\,k}{}_l
\ed z^{\,i}{}_j + (t-1) z^{\,i}{}_j \ed z^{\,k}{}_l + \hat{\sigma} z^{\,i}{}_j
z^{\,k}{}_l \Trq \theta \\[1ex]
& & + t (q-\frac{1}{q}) (z^{\,i}{}_l \ed z^{\,k}{}_j + z^{\,k}{}_j \ed
z^{\,i}{}_l) + t(q-\frac{1}{q})^2 z^{\,i}{}_j \ed z^{\,k}{}_l \\[1ex]
& & - \frac{1}{q^2} \hat{\beta} (z^{\,i}{}_j z^{\,k}{}_l - q z^{\,i}{}_l
z^{\,k}{}_j) \Trq
\theta & i < k,j < l \\[1ex]
(\ed z^{\,i}{}_j) z^{\,k}{}_l & = & t z^{\,k}{}_l
\ed z^{\,i}{}_j + (t-1) z^{\,i}{}_j \ed z^{\,k}{}_l + \hat{\sigma} z^{\,i}{}_j
z^{\,k}{}_l \Trq \theta \\[1ex]
& & + t(q-\frac{1}{q}) z^{\,i}{}_l \ed z^{\,k}{}_j + \frac{1}{q} \hat{\beta}
(z^{\,i}{}_l z^{\,k}{}_j - q z^{\,i}{}_j z^{\,k}{}_l) \Trq \theta &
i < k,j > l \\[1ex]
\end{array}
\end{equation}
with $t \in \Cx \setminus \{0\},\ (q^6+q^2+1)t-(q^4+q^2+1) \neq 0$.
We have introduced the abbreviations
\setlength{\mathindent}{0.5cm}
\begin{eqnarray}
   \Trq \theta & = & q^4 \theta^{\,1}{}_1 + q^2 \theta^{\,2}{}_2 +
\theta^{\,3}{}_3 \ ,\\[1ex]
   \sigma & = & \frac{(q^2-t)(t-1)}{q^4(q^2+1)(t-1) -q^2+t} \ , \\[1ex]
   \beta & = & \frac{t(q^2-1)(t-1)}{q^4(q^2+1)(t-1) -q^2+t} \ , \\[1ex]
   \hat{\sigma} & = & -\frac{(q^2t-1)(t-1)}{q^4(q^2t-1) + (q^2+1)(t-1)}
\ , \\[1ex]
   \hat{\beta} & = & -\frac{t(q^2-1)(t-1)}{q^4(q^2t-1) + (q^2+1)(t-1)}
\ .
\end{eqnarray}
The missing relations can be derived in both cases by using the
Leibniz rule and the relations \th{(\ref{explRZZ})}.
\end{cor}
\setlength{\mathindent}{1cm}

{\bf Remark.} For the special case $t=1$\ the formulas (\ref{dzz1})
and (\ref{dzz2}) simplify drastically. They can be written in
compact form
\begin{eqnarray} \label{dzz1t=1mat}
   \ed Z_1 Z_2 & = & R_{12}^{-1} Z_2 \ed Z_1 R_{21}^{-1} \ , \\[1ex]
   \label{dzz2t=1mat}
   \ed Z_1 Z_2 & = & R_{21} Z_2 \ed Z_1 R_{12} \ ,
\end{eqnarray}
for $\nu = 1$\ and $\nu = 2$,
respectively, and define bicovariant differential calculi for
arbitrary $n$. These relations were first found by Maltsiniotis
\cite{Mal90} and independently by Manin \cite{Ma91}. They
investigated differential calculi on multi-parameter deformations of
GL$(n)$ that are induced by calculi on the corresponding quantum plane.
In $R$-matrix form (\ref{dzz1t=1mat}) and (\ref{dzz2t=1mat}) appeared
in \cite{Schirr92} and \cite{Sud92,Sud93} and were studied in detail
in \cite{SWZ92} (see also \cite{Zu92}).

The bi-invariant element $\Trq \, \theta$ plays a particular role.
Acting with it on $\mca$ by taking the commutator
$[\Trq \, \theta, a]$ ($a \in \mca$) defines a derivation from
$\mca$ into the space of 1-forms. It turns out that this derivation
coincides with $\ed$ up to a normalization factor.

\begin{prop} For all first order differential calculi
$(\Gamma_{\nu}(t), \ed)$\ on \th{GL$_q(3,\Cx)$}\ the differential $\ed$
is an inner derivation:
\begin{equation} \label{innerder}
   \ed a = \frac{1}{\cal N} \, [\Trq \theta, a]
\end{equation}
where
\begin{equation}
   {\cal N} = \left\{ \begin{array}{c@{\quad \mbox{for} \quad}l}
               \frac{1}{q^2} (q^4(q^2+1)(t-1)-q^2+t) & \nu=1 \\[1ex]
               q^4(q^2t-1)+(q^2+1)(t-1) & \nu=2 \end{array} \right.
\end{equation}
\end{prop}

\subsection{R-matrix formulation}  \label{subRmat}

The commutation relations of GL$_q(3,\Cx)$\ can be written in the
compact form (\ref{RZZ}) using the $R$-matrix (\ref{R-mat}). Now
the question arises whether also the bimodule structure of
$\Gamma_{\nu}(t)$ can be compactly expressed in such a way. Indeed,
this can be achieved by using a convenient basis of $\Gal$. It is
related to a procedure proposed by Jur\v{c}o \cite{Ju91} to construct
bicovariant differential calculi on certain (classes of)
quantum groups. The latter can be applied to the case of GL$_q(n,\Cx)$
for arbitrary dimension $n$. The construction is based on a further
result of Woronowicz \cite{Wo89} which we recall next.

Given a family of functionals $f = (\zwi f I J)_{I,J \in \cal I}$\ and a
family of algebra elements $M = (\zwia M I J)_{I,J \in \cal I}$\
satisfying (\ref{re-prop}), (\ref{re-prop1}), (\ref{ad-re}),
(\ref{ad-re1}) and the compatibility condition (\ref{bico-con}) one
can endow the free left \ca-module \Ga generated by $\{\omega^{\,I}\}_{I
\in \cal I}$\ with the structure of a bicovariant bimodule: One
regards $\{\omega^{\,I}\}$\ as left-invariant elements forming a basis of
\Gal \ and defines the right multiplication by (\ref{comw}) and the right
coaction by (\ref{rcoom}).

It is easy to see that $M = Z$ and $M = S(Z)^t$ are possible choices
for $M$ ($^t$\ denotes ordinary matrix transposition). The appropriate
functionals are the generators $L^{\pm} = (\zwi{\lpm}{i}{j})_{1\leq
i,j \leq n}$ of the algebra of regular functionals on \GL.
They are defined by \cite{FRT}
\begin{equation} \label{def-l-fun}
   \begin{array}{rcl}
   \langle \zwi{\lpm}{i}{j}, \zz k l \rangle & = & \vii{R^{\pm}{}} i j
k l \\[1ex]
   \langle \zwi{\lpm}{i}{j}, \un \rangle & = & \kro i j  \\[1ex]
   \langle \zwi{\lpm}{i}{j}, ab \rangle & = & \langle \zwi{\lpm}{i}{k}, a
\rangle \langle \zwi{\lpm}{k}{j}, b \rangle
   \end{array}
\end{equation}
for all $a,b \in \mca$\ where we denote the evaluation $\ell(a)$\ by
$\langle \ell, a \rangle$\ and use the abbreviations
\begin{equation}
   R^+ = c^+PRP, \qquad R^- = c^-R^{-1}.
\end{equation}
Here $P$\ is the permutation matrix $\vii P i j k l = \kro i l \kro k
j$\ and $c^+, c^-$\ are complex constants $\neq 0$. The quantum
Yang-Baxter equation (\ref{QYBE}) assures the compatibility of
(\ref{def-l-fun}) with the relations (\ref{RZZ}). The dual of \ca
denoted by $\mca'$ has a natural multiplication structure given by
the convolution product
\begin{displaymath}
   \langle f \ast g, a \rangle = \langle f \otimes g, \Delta a
\rangle \ , \quad a\in \mca, f,g \in \mca'\ ,
\end{displaymath}
and contains $\varepsilon$ as unit element. One regards the
subalgebra $\cal U$ of \ca generated by $\zwi{\lpm} i j$ (and two
further functionals $\lpm$ playing a similar role as the
inverse of the quantum determinant in the construction of the
Hopf-algebra \ca). $\cal U$ can be endowed with the structure of a
Hopf-algebra in a natural way (cf \cite{FRT,SW94}). In
particular, one obtains for the antipode $S'$
\begin{equation}
   \langle S'(L^{\pm}), Z \rangle = \langle L^{\pm}, S(Z) \rangle =
(R^{\pm})^{-1} \ .
\end{equation}

It turns out that in the case of $M = Z$ the choice
$f = S'(L^{\pm})^t$ fulfils all requirements mentioned above. The
condition (\ref{bico-con}) is checked on the generators $a = \zz i j$
with the help of the basic relations (\ref{RZZ}). For $M = S(Z)^t$ one
sets $f = L^{\pm}$. However, in these cases one is led to bicovariant
bimodules of dimension $n$. To build up an $n^2$-dimensional bimodule
as a candidate for a differential calculus on GL$(n,\Cx)$ tensor products
of two $n$-dimensional bimodules can be used. Out of the various
possibilities \cite{Ju91} we choose
\begin{equation} \label{tpfunct}
   \begin{array}{r@{\: = \:}c@{\: = \:}l}
   \zwi M I J & \viia M i j k l & \zz i k S(\zz l j) \ , \\[1ex]
   \label{sll}
   \zwia f I J & \viib f i j k l & S'(\zwi{\lpm}{k}{i}) \ast
\zwi{\lmp}{j}{l} \ .
   \end{array}
\end{equation}
The commutation relations of the bimodule generators $\zwia{\omega} i
j$\ and the algebra generators $\zz k l$\ are for the choice of upper
signs in (\ref{sll})
\begin{equation} \label{Rmatcom1}
   \zwia{\omega} i j  \zz k l = t \zz k d
\vii{(R^{-1})} d e a i \vii{(R^{-1})} j b e l \zwia{\omega} a b
     \qquad (t = c^-/c^+ \neq 0)
\end{equation}
and in the case of lower signs
\begin{equation} \label{Rmatcom2}
   \zwia{\omega} i j  \zz k l = t \zz k d
\vii{R} a i d e \vii{R} e l j b \zwia{\omega} a b
     \qquad (t = c^+/c^- \neq 0)   \, .
\end{equation}
These have the desired simple form.

To introduce a differential operator $\ed$\ one uses (in both cases)
the bi-invariant element $\mbox{Tr}\, \omega = \sum_i \zwia{\omega} i i$.
$\ed a$\ is defined for all $a \in \mca$\ as
\begin{equation}
   \ed a = \frac{1}{q-q^{-1}} \, [\mbox{Tr} \, \omega, a] \ .
\end{equation}
$\ed$ satisfies the Leibniz rule and using the bi-invariance of
$\mbox{Tr} \, \omega$\ one can verify (\ref{lcod}) and (\ref{rcod}).
Now it is possible to calculate the relation between $\zwia{\omega} i j$
and the Maurer-Cartan 1-forms defined in (\ref{MCforms}). One
obtains\footnote{Here the double index $^i{}_j$\ determines the row,
$^k{}_l$\ the column of the matrix $U$.}
\begin{equation}  \label{trafothom}
   \tz i j = \vii U i j k l  \, \zwia{\omega} k l
\end{equation}
where the complex matrix $U \in M(n^2,\Cx)$\ is given by
\begin{eqnarray} \label{U1}
   \vii U i j k l & = & \frac{1}{q-q^{-1}}(t \vii{(R^{-1})} i a k b
\vii{(R^{-1})} b l a j - \kro i j \kro k l) \\ \label{U2}
   \vii U i j k l & = & \frac{1}{q-q^{-1}}(t \vii{R} k a i b
\vii{R} b j a l - \kro i j \kro k l)
\end{eqnarray}
in the first and second case, respectively. \Ga is generated by $\ed
\zz i j $\ as a left \ca-module\footnote{Recall the additional
assumption at the beginning of this section.} if and only if $U$ is
invertible. This leads to additional restrictions on $t$, in the case
$n=3$ these are
\begin{displaymath}
   \begin{array}{r@{\, \neq \,}l@{\quad \mbox{for}\ \nu\, = \,}l}
   (q^6+q^4+1)t-(q^6+q^4+q^2) & 0 & 1\ , \\[1ex]
   (q^6+q^2+1)t-(q^4+q^2+1) & 0 & 2\ .
   \end{array}
\end{displaymath}
Using the transformation (\ref{trafothom}), the relations
(\ref{Rmatcom1}) and (\ref{Rmatcom2}) lead to commutation relations
of Maurer-Cartan 1-forms and algebra generators which agree with
those found in \ref{subres} for the differential calculi
$\Gamma_{\nu}(t)$.

\begin{prop} Let $q \in \Cx \setminus \{0,\pm 1,\pm i\}$. For
every bicovariant differential calculus on \th{\GL} there is a basis
of \Gal\ such that the commutation relations \th{(\ref{comw})} can be
expressed in terms of the $R$-matrix as follows. For the calculi
$\Gamma_1(t)$\ this basis is given by \th{(\ref{trafothom})} and
\th{(\ref{U1})} and leads to relations \th{(\ref{Rmatcom1})}. In the case of
$\Gamma_2(t)$\ the relations \th{(\ref{Rmatcom2})} are obtained with the
transformation given by \th{(\ref{trafothom})} and \th{(\ref{U2})}.
\end{prop}

{\bf Remark.} The procedure outlined above has been used in several
papers to construct examples of bicovariant differential calculi on
quantum groups. The calculi $\Gamma_1(t)$ are discussed in
\cite{CA93,CA92} for GL$_q(2,\Cx)$\ and GL$_q(3,\Cx)$.\footnote{The
statement in \cite{CA92} that the additional parameter is inessential
is incorrect as we have shown.} In \cite{SW94} the calculi
$\Gamma_2(t)$ were given for GL$_q(n,\Cx)$.
It is interesting that this procedure already exhausts the possible
bicovariant differential calculi in the case of \GL. For
GL$_{p,q}(2,\Cx)$ this has been shown in \cite{MH93}. In that case there
is only one family of calculi.

\section{Induced calculi on SL$_q(3,\Cx)$\ and real forms}
\setcounter{equation}{0}

With the complete collection of bicovariant differential calculi on
GL$_q(3,\Cx)$ at hand one can proceed to investigate the induced
calculi on quantum subgroups.
Those are obtained by imposing additional relations on \ca or by
introducing an involution (a $\ast$-structure).

\subsection{SL$_q(3,\Cx)$ as quantum subgroup of GL$_q(3,\Cx)$}

The quantum group \SL\ is obtained from \GL\ by adding the
unimodularity condition
\begin{equation} \label{unimod}
   \de = \un  \ .
\end{equation}
This is consistent with the Hopf algebra structure of \GL. As an
immediate consequence we have
\begin{equation} \label{ddet=0}
   \ed \de = 0
\end{equation}
for a differential calculus over \SL. We determine all bicovariant
differential calculi on \SL\ which are `induced' by a differential
calculus on \GL. These are all calculi on \GL\ that are consistent
with the additional conditions (\ref{unimod}) and (\ref{ddet=0}).
Acting with ${\cal F}$ on (\ref{unimod}) leads to
\begin{equation} \label{tqdep}
   t^3 q^{\mp 2} = 1
\end{equation}
with $-$ for the first and $+$ for the second family of calculi.
Calculation of $\ed \de$ leads to
\begin{eqnarray}
   \ed \de & = & \frac{t^3-q^2}{q^4(q^2+1)(t-1) -q^2+t} \de \, \Trq
\theta \\[1ex]
   \ed \de & = & \frac{q^2t^3-1}{q^4(q^2t-1) + (q^2+1)(t-1)} \de \, \Trq
\theta
\end{eqnarray}
for the first and second case, respectively.
All this can be summarized as follows.

\begin{prop} Let $q \in \Cx \setminus \{0,\pm 1,\pm i\}$. In order to
obtain bicovariant differential calculi on \th{\SL} from
\th{(\ref{dzz1})} and \th{(\ref{dzz2})} one has to set $t^3 = q^2$
and $t^3 = q^{-2}$, respectively. Hereby solutions of
\th{(\ref{tqdep})} with
\begin{displaymath}
   \begin{array}{r@{\, = \,}l@{\quad \mbox{for}\ \nu\, = \,}l}
   (q^6+q^4+1)t-(q^6+q^4+q^2) & 0 & 1\ , \\[1ex]
   (q^6+q^2+1)t-(q^4+q^2+1) & 0 & 2\ .
   \end{array}
\end{displaymath}
have to be excluded.
Hence, for generic $q$ there are six bicovariant differential calculi
on \th{\SL}.
\end{prop}

{\bf Remark.} Though (\ref{unimod}) constrains the $\zz i j$, their
differentials remain independent with regard to the left module
structure. It is impossible, for example, to express $\ed z^9 \ $as
$\ed z^9 = a_I \ed z^I,\ I = 1, \ldots ,8$. This means that all
bicovariant differential calculi on \SL\ given above have nine
independent 1-forms. Indeed, as was shown in \cite{SchSch94} the
dimension of the space of 1-forms on SL$(n,\Cx)$ is fixed to be 1 or
$n^2$ if bicovariance is assumed.

\subsection{Real forms of \GL and \SL}

To obtain real forms of the quantum group \GL\ one has to endow the
underlying Hopf-algebra with a $\ast$-structure, i.e. a linear map
$\ast : \mca \longrightarrow \mca$\ with
\begin{equation}
   \begin{array}{c@{\;=\;}c@{\qquad}c@{\;=\;}c}
   (ab)^{\ast} & b^{\ast}a^{\ast} & \Delta(a^{\ast}) &
\Delta(a)^{\ast} \\[1ex]
   (\lambda a)^{\ast} & \overline{\lambda} a^{\ast} & \epsilon(a^{\ast}) &
\overline{\epsilon(a)} \\[1ex]
   (a^{\ast})^{\ast} & a & S(S(a)^{\ast})^{\ast} & a
   \end{array}
\end{equation}
for all $a,b \in \mca,\ \lambda \in \Cx$. Usually there are different
choices for such a $\ast$-structure. We consider two of them
\cite{FRT}:

(1) The quantum group GL$_q(3,\Rl)$ is obtained by setting
\begin{equation} \label{real*}
   Z^{\ast} = Z\ , \qquad (\ide)^{\ast} = \ide\ .
\end{equation}
The action of $^{\ast}$ is extended to the whole algebra \ca as an
antihomomorphism. For this to be welldefined, i.e. to be compatible
with the relations (\ref{RZZ}), one has to demand $|q| = 1$.

(2) Analogously one introduces for $q \in \Rl$\ the notion of
hermitian conjugation by
\begin{equation} \label{unitary*}
   Z^{\ast} = S(Z)^t, \qquad (\ide)^{\ast} = \de\ .
\end{equation}
and obtains the quantum unitary group U$_q(3)$.

By imposing additionally the unimodularity condition (\ref{unimod})
one is led to the quantum groups SL$_q(3,\Rl) \ (|q|=1)$\ and SU$_q(3)
\ (q \in \Rl)$, respectively.

A bicovariant differential calculus on a $\ast$-Hopf algebra should
admit an extension of the $\ast$-operation to the space of 1-forms
\Ga in such a way that (cf \cite{Wo89})
\begin{equation} \label{starstru}
   \begin{array}{c@{\;=\;}c}
   (a \varrho)^{\ast} & \varrho^{\ast} a^{\ast}, \\[1ex]
   (\varrho a)^{\ast} & a^{\ast} \varrho^{\ast}, \\[1ex]
   (\ed a)^{\ast} & \ed (a^{\ast}).
   \end{array}
\end{equation}
As a consequence one has the compatibility of the $\ast$-structure
with the left and right coaction of \ca on \Ga:
\begin{equation}
   \begin{array}{c@{\;=\;}c}
   \lco(\varrho^{\ast}) & \lco(\varrho)^{\ast},  \\[1ex]
   \rco(\varrho^{\ast}) & \rco(\varrho)^{\ast}.  \\[1ex]
   \end{array}
\end{equation}

Given a $\ast$-structure as well as a bicovariant differential
calculus on \GL, there is at most one $\ast$-structure on $\tz i j$\
that fulfils all requirements (\ref{starstru}). We discuss the
results in the case of the two examples above.

(1) In the case of (\ref{real*}) one deduces with the help of
(\ref{starstru}) the formula
\begin{equation}
   (\tz i j)^{\ast} = q^{2(n-i)} \viia f n j k l (S(\zz i n)) \tz k l
\ .
\end{equation}
For the calculi $\Gamma_1(t)$\ this reads explicitly
\begin{equation}
   \begin{array}{rcl}
   (\tz 1 1)^{\ast} & = & \frac{q^6}{t^2} \tz 1 1 +
\frac{q^2}{t^2N}(t-1)(t-q^6) \Trq \, \theta \\[1ex]
   (\tz 1 i)^{\ast} & = & \frac{q^5}{t^2} \tz 1 i \qquad \mbox{for}\ i =
2,3 \\[1ex]
   (\tz 2 2)^{\ast} & = & \frac{q^2}{t^2} \tz 2 2 +
\frac{q^2}{t^2N}((t-1)(1-q^6)+t(q^2+\frac{1}{q^2}-2)) \tz 3 3 \\[1ex]
& & + \frac{q}{t^2N}(q^2t-1)(t-q^2) \Trq \, \theta \\[1ex]
   (\tz 2 i)^{\ast} & = & \frac{q^3}{t^2} \tz 2 i \qquad \mbox{for}\ i =
1,3 \\[1ex]
   (\tz 3 3)^{\ast} & = & \frac{1}{t^2} \tz 3 3 +
\frac{1}{t^2N}(q^2t-1)(t-q^2) \Trq \, \theta \\[1ex]
   (\tz 3 i)^{\ast} & = & \frac{q}{t^2} \tz 3 i \qquad \mbox{for}\ i =
1,2
   \end{array}
\end{equation}
with $N = q^4(q^2+1)(t-1)-q^2+t$. In the case of $\Gamma_2(t)$\ we
have similarly
\begin{equation}
   \begin{array}{rcl}
   (\tz 1 1)^{\ast} & = & \frac{1}{t^2} \tz 1 1 +
\frac{1}{q^2t^2N}(q^2t-1)(t-q^2) \Trq \, \theta \\[1ex]
   (\tz 1 i)^{\ast} & = & \frac{1}{qt^2} \tz 1 i \qquad \mbox{for}\ i =
2,3 \\[1ex]
   (\tz 2 2)^{\ast} & = & \frac{1}{q^4t^2} \tz 2 2 +
\frac{1}{q^6t^2N}(q^4t(q^2-1)-q^4(t-q^2)+t-1) \tz 3 3 \\[1ex]
& & + \frac{1}{q^6t^2N}(q^6t-1)(t-1) \Trq \, \theta \\[1ex]
   (\tz 2 i)^{\ast} & = & \frac{1}{q^3t^2} \tz 2 i \qquad \mbox{for}\ i =
1,3 \\[1ex]
   (\tz 3 3)^{\ast} & = & \frac{1}{q^6t^2} \tz 3 3 +
\frac{1}{q^6t^2N}(q^6t-1)(t-1) \Trq \, \theta \\[1ex]
   (\tz 3 i)^{\ast} & = & \frac{1}{q^5t^2} \tz 3 i \qquad \mbox{for}\ i =
1,2
   \end{array}
\end{equation}
with $N = q^4(q^2t-1) + (q^2+1)(t-1)$. For $\ast$\ to be an
involution it is necessary to require $|t| = 1$. If $\varrho$\ is an
arbitrary element of \Ga with $\varrho = a_I \theta^I$\ we set
$\varrho^{\ast} = (\theta^I)^{\ast}(a_I)^{\ast}$. Then we can proof
that (\ref{starstru}) holds indeed using the commutation relations
(\ref{comth}) and the property (\ref{innerder}) observing that
\begin{equation} \label{*derivator}
   (\frac{1}{\cal N} \Trq \, \theta)^{\ast} = - \frac{1}{\cal N}
          \Trq \, \theta \, .
\end{equation}

\begin{prop}
   Let $q \in \{w \in \Cx \mid |w| = 1 \} \setminus \{\pm 1, \pm i
\}$. Then all bicovariant $\ast$-calculi on \th{GL$_q(3,\Rl)$} are
given by \th{(\ref{dzz1})} and \th{(\ref{dzz2})} with the restriction $|t|=1$
in both cases. All 6 calculi on \th{\SL} found for generic $q$ are
$\ast$-calculi.
\end{prop}

(2) For U$_q(3)$\ the only $\ast$-structure on $\Gamma_{\nu}(t)$
 is given by
\begin{equation} \label{unitary*th}
   (\theta^{\,i}{}_j)^{\ast} = - \theta^{\,j}{}_i \ .
\end{equation}
Using (\ref{comth}) one proves that $(\varrho a)^{\ast} = a^{\ast}
\varrho^{\ast}$ holds if and only if $t$ is real. Again,
(\ref{*derivator}) holds as a consequence of (\ref{unitary*th}) and the
reality of $t$. Hence $(\ed a)^{\ast} = \ed (a^{\ast})$.

\begin{prop} Let $q \in \Rl \setminus \{0,\pm 1 \}$. All
bicovariant $\ast$-calculi on \th{U$_q(3)$}\ are given by \th{(\ref{dzz1})}
or \th{(\ref{dzz2})} with $t \in \Rl$. On \th{SU$_q(3)$} these induce two
bicovariant $\ast$-calculi corresponding to the real solutions of
$t^3 = q^{\pm 2}$.
\end{prop}

{\bf Remark.} On SU$_q(2)$ one recovers the $4D_{\pm}$ calculi
\cite{Wo89}. The uniqueness of the latter has been shown in
\cite{Stach92}. In \cite{CWSWW91} and \cite{SuWa93} examples of bicovariant
differential calculi on SU$_q(n)$ for arbitrary $n$ are given with the help
of the constructive procedure outlined in 4.4.
In \cite{SuWa93} the $n$ calculi corresponding to the choice of lower
signs in (\ref{tpfunct}) and the parameter values $t^n = q^{-2}$ are
discussed. The authors claim that all these calculi are $*$-calculi. This is
not true, however, for $t \not\in \Rl$.

\section{The classical limit}
\setcounter{equation}{0}

It is interesting to investigate the behavior of the differential
calculi on \GL\ and \SL\ in the limit $q \rightarrow 1$. One might
expect the classical calculus to emerge. However, the formulas
obtained for $q \to 1$\ depend on the way in which the limit is
performed.

In the case of \GL\ the additional parameter $t$ may depend on $q$
but need not. If $t$ and $q$ are regarded as independent, we obtain
with
\begin{equation}
\begin{array}{ccccc}
   \lim\limits_{q\to 1} \sigma & = & \lim\limits_{q\to 1} \hat{\sigma} & = &
\frac{1-t}{3}
\\[1ex]
   \lim\limits_{q\to 1} \beta & = & \lim\limits_{q\to 1} \hat{\beta} & = & 0
\end{array}
\end{equation}
from (\ref{dzz1}) and (\ref{dzz2}) a one-parameter family of calculi on
GL$(3,\Cx)$:
\begin{equation}  \label{dzzq=1}
\begin{array}{rclc}

(\ed z^{\,i}{}_j) z^{\,i}{}_j & = & (2t-1) z^{\,i}{}_j
\ed z^{\,i}{}_j + \frac{1-t}{3} z^{\,i}{}_j z^{\,i}{}_j \Trq \theta &  \\[1ex]
(\ed z^{\,i}{}_j) z^{\,i}{}_l & = & t z^{\,i}{}_l
\ed z^{\,i}{}_j + (t-1) z^{\,i}{}_j \ed z^{\,i}{}_l + \frac{1-t}{3} z^{\,i}{}_j
z^{\,i}{}_l \Trq \theta & j < l \\[1ex]
(\ed z^{\,i}{}_j) z^{\,k}{}_j & = & t z^{\,k}{}_j
\ed z^{\,i}{}_j + (t-1) z^{\,i}{}_j \ed z^{\,k}{}_j + \frac{1-t}{3} z^{\,i}{}_j
z^{\,k}{}_j \Trq \theta & i < k \\[1ex]
(\ed z^{\,i}{}_j) z^{\,k}{}_l & = & t z^{\,k}{}_l
\ed z^{\,i}{}_j + (t-1) z^{\,i}{}_j \ed z^{\,k}{}_l + \frac{1-t}{3} z^{\,i}{}_j
z^{\,k}{}_l \Trq \theta & i < k,j < l \\[1ex]
(\ed z^{\,i}{}_j) z^{\,k}{}_l & = & t z^{\,k}{}_l
\ed z^{\,i}{}_j + (t-1) z^{\,i}{}_j \ed z^{\,k}{}_l + \frac{1-t}{3} z^{\,i}{}_j
z^{\,k}{}_l \Trq \theta & i < k,j > l \\[1ex]
\end{array}
\end{equation}
For $t \rightarrow 1 $ one recovers the classical calculus where
$\left[\ed z^{\,i}{}_j, z^{\,k}{}_l \right] = 0 \ \forall \, i,j,k,l$.
We can obtain calculi on SL$(3,\Cx)$\ from (\ref{dzzq=1}) by imposing
(\ref{unimod}) which fixes $t$\ to be a solution of $t^3 = 1$.
Apart from the classical calculus one is led in this way to two
non-classical calculi corresponding to the two primitive third roots
of unity.

In the case of \SL\ we meet with a different situation. Since $t \
\mbox{and} \ q$ are related by (\ref{tqdep}), $t$ is determined for $q
\rightarrow 1$ up to the fact that a cubic equation for $t$ has
three solutions in the complex plane. For $t \to \xi$\ and $t \to
\xi^2$ with $\xi = e^{(2\pi i/3)}$ one finds the same result as by
setting $t = \xi$\ or $t = \xi^2$ in (\ref{dzzq=1}). Here we investigate
the case $t = q^{\pm 2/3} \rightarrow 1$ in some more detail:
\begin{equation}
\begin{array}{ccccc}
   \lim\limits_{q\to 1} \sigma & = & \lim\limits_{q\to 1} \hat{\sigma} & = &
\frac{1}{6} \ ,
\\[1ex]
   \lim\limits_{q\to 1} \beta & = & \lim\limits_{q\to 1} \hat{\beta} & = &
\frac{1}{4}  \ .
\end{array}
\end{equation}
This leads us in both cases (\ref{dzz1}), (\ref{dzz2}) to the following
structure:
\begin{equation} \label{taustru4}
   \left[\ed z^{\,i}{}_j, z^{\,k}{}_l \right] =
 \gamma^{\,i}{}_j{}^{\,k}{}_l \, \tau
\end{equation}
with the abbreviations
\begin{equation}  \label{taugamma}
   \begin{array}{r@{\: = \:}l}
   \tau & \frac{3}{2} \mbox{Tr} \,\theta  =  \frac{3}{2} (\theta^{\,1}{}_1
   + \theta^{\,2}{}_2 + \theta^{\,3}{}_3) \ ,\\[1ex]
   \gamma^{\,i}{}_j{}^{\,k}{}_l & \frac{1}{6}(
   z^{\,i}{}_l z^{\,k}{}_j - \frac{1}{3} z^{\,i}{}_j z^{\,k}{}_l)\ .
   \end{array}
\end{equation}
Using composite indices we have
\begin{equation} \label{taustru}
   \left[\ed z^I, z^J \right] = \gamma^{IJ} \, \tau \ , \quad \tau =
   \tau_J \, \ed z^J \ .
\end{equation}
The symmetric matrix $\gamma$ is degenerate, i.e. $\det \gamma = 0$,
and satisfies $\gamma^{IJ} \tau_J = 0$. One of the `coordinates'
$z^I$ is redundant because of the constraint $\de = \un$. We can
eliminate e.g. $z^9$ in a certain coordinate patch, where $z^1 z^5
- z^2 z^4 \neq 0$. If we consider in (\ref{taustru}) only indices
$I,J = 1, \ldots , 8$, then we obtain a nondegenerate part of
$\gamma$,
\begin{equation}
   g = (\gamma^{IJ})_{1 \leq I,J \leq 8}
\end{equation}
with $\det g = -(z^1 z^5-z^2 z^4)^2/(3\cdot 6^8) \neq 0$. The 1-form
$\tau$ is still independent of the 1-forms $\ed z^I, I = 1,\ldots,8$. In
particular, $\mbox{Tr} \,\theta$ does not vanish in the classical
limit.

The matrix $g^{-1}$ gives rise to a metric
\begin{equation}
  B = g_{IJ} \, \ed z^I \otimes \ed z^J
\end{equation}
on SL$(3,\Cx)$ (where we set $g_{IK}g^{KJ} = \delta_{I}^{J}$)
which turns out to be the Cartan-Killing metric.
In order to prove this we first introduce the Maurer-Cartan 1-forms
$\htz i j$ corresponding to the ordinary differential calculus on
SL$(3,\Cx)$. They are given by $\hat{\theta}= Z^{-1}\ed Z$ and obey
$\mbox{Tr}\, \hat{\theta}=0$. In terms of the basis $\{\hat{\theta}^I \mid
I=1,\ldots,8\}$ of the space of 1-forms on SL$(3,\Cx)$ we have
\begin{equation}
   B = \hat{g}_{IJ}\, \hat{\theta}^I \otimes \hat{\theta}^J
\end{equation}
with the coefficient matrix
\begin{equation} \label{gmat}
   (\hat{g}_{IJ}) =  6\, \left( \begin{array}{cccccccc}
2 & 0 & 0 & 0 & 1 & 0 & 0 & 0 \\
0 & 0 & 0 & 1 & 0 & 0 & 0 & 0 \\
0 & 0 & 0 & 0 & 0 & 0 & 1 & 0 \\
0 & 1 & 0 & 0 & 0 & 0 & 0 & 0 \\
1 & 0 & 0 & 0 & 2 & 0 & 0 & 0 \\
0 & 0 & 0 & 0 & 0 & 0 & 0 & 1 \\
0 & 0 & 1 & 0 & 0 & 0 & 0 & 0 \\
0 & 0 & 0 & 0 & 0 & 1 & 0 & 0
\end{array} \right).
\end{equation}
On the other hand, the Cartan-Killing metric $\kappa$ on SL$(3,\Cx)$
can be written as \cite{Hel78}
\begin{equation} \label{caki}
   \kappa(\tilde{X},\tilde{Y}) = 6 \, \mbox{Tr}(XY)
\end{equation}
where $\tilde{X}$ and $\tilde{Y}$ are the leftinvariant vector fields
generated by $X,Y \in s\ell(3,\Cx)$. The basis $\{X_I \mid I=1,\ldots,8\}$
of $s\ell(3,\Cx)$ that generates vector fields dual to $\{\hat{\theta}^I
\mid I=1,\dots,8\}$ is given by
\begin{displaymath}
   \begin{array}{r@{\;=\;}l@{\quad for\ }l}
      \zwia X i j & \zwia e i j & i\neq j \, ,\\[1ex]
      \zwia X i i & \zwia e i i - \zwia e 3 3 & i = 1,2\, .
   \end{array}
\end{displaymath}
The matrices $\zwia e i j$ are defined by
$(\zwia e i j)^{\,k}{}_l = \kro k i \kro l j$.
Using (\ref{caki}) and (\ref{gmat}) one easily obtains
\begin{displaymath}
   \begin{array}{r@{\;=\;}l}
      \kappa & \kappa(\zwia{\tilde{X}} i j,\zwia{\tilde{X}} k l)\,
      \htz i j \otimes \htz k l \\[1ex]
& 6(\kro l i \kro j k + \kro j i \kro l k)\, \htz i j \otimes \htz k
l \\[1ex]
& \hat{g}_i{}^j{}_k{}^l \, \htz i j \otimes \htz k l\ .
   \end{array}
\end{displaymath}
Consequently, $B$ equals the Cartan-Killing metric, which is
bi-invariant and has signature (5,3).
The bicovariant differential calculi on \SL\ are compatible with the
`reality conditions' $(z^I)^{\ast} = z^I$, so that we obtain the same
result for SL$_q(3,\Rl), q \rightarrow 1$. Then $\gamma$\ and $\tau$\
form a (generalized) `Galilei structure'
on the group manifold SL$(3,\Rl)$. A corresponding result
for SL$_q(2,\Rl)$ was obtained in \cite{MHR93} (see also \cite{DM93}).

\section{Conclusions}
\setcounter{equation}{0}

The way we obtained our results is not restricted to specific values
of $n$, in principle. However, even for $n=3$ computations are
lengthy and tedious.
We proved that for GL$_q(3,\Cx)$ there are only two one-parameter
families of bicovariant differential calculi which both can be obtained
by Jur\v{c}o's method described in \ref{subRmat}. Out of these (for
generic $q$) there are six calculi that are consistent with the condition
of unimodularity. In this way one is led to all 9-dimensional
bicovariant differential calculi on \SL.

There have been attempts to construct bicovariant differential calculi on
SL$_q(n,\Cx)$\ with an $(n^2-1)$-dimensional space of 1-forms
\cite{IP94,FP94} that are also bicovariant. This can only be achieved if
one allows a deformation of the ordinary Leibniz rule for the
exterior differential. The great advantage of keeping the latter is,
however, its universality and simplicity.

On the other hand following the path outlined above one arrives at an
interesting deformation of the ordinary calculus on SL$(n,\Rl)$ that
was discussed in a more general setting in \cite{DM93,DM92}. There
it has been pointed out that similar structures can be found in the
It\^{o} calculus of stochastic differentials. Also, relations to proper
time formulations of quantum theories have been established. All this
hints at a possible physical relevance of the structure (\ref{taustru}).
For SL$(n,\Rl)$ the natural
group metric enters this formula. This motivates further investigations
concerning a suitable generalization to the case $q \neq 1$. It seems
to be reasonable that a candidate for a quantum group metric can be
obtained this way. This would be a crucial step in gaining more insight
into the geometry of a quantum group and could pave the way to a
formulation of Kaluza-Klein theories using quantum groups as internal
spaces.

After completion of this work we received a preprint \cite{Sch94} in
which a complete classification of bicovariant differential calculi
on GL$_q(n,\Cx)$ for arbitrary $n$ is reported. The methods used
there are different from ours. Our discussion of the case $n=3$ is
more detailed and clarifies the relation to work by other authors. In
particular, we have presented explicit formulas for the commutation
relations of the algebra generators $\zz i j$ and their
differentials. We have discussed calculi on real forms of \GL\ and
considered the classical limit in some detail. Of most interest
hereby is the geometric structure which arises in the classical limit
of a bicovariant differential calculus on SL$(n, \Cx)$.

\vspace{0.5cm}
{\bf Acknowledgments.} I would like to thank F.\ M\"uller-Hoissen for
numerous fruitful discussions and his permanent encouragement as well
as for many valuable comments on a preliminary version of this
manuscript. I am also grateful to H.\ Dathe for a helpful remark
on the structure of some {\sc Reduce} programs.
\pagebreak

\end{document}